\documentclass[CMYK,NoSecthm]{jcomsec_author}  
\usepackage[linkcolor=black,citecolor=black,colorlinks=false,bookmarksnumbered]{hyperref} 
\usepackage{multicol}
\usepackage{enumerate}
\usepackage{amssymb}
\usepackage{multirow}
\usepackage{amsmath}
\usepackage{graphicx}
\usepackage{subcaption}
\usepackage{colortbl}
\usepackage{color}
\usepackage{wrapfig}
\usepackage[numbers,sort&compress]{natbib}
\usepackage[all]{hypcap}
\usepackage{array}
\usepackage{microtype}

\DisableLigatures[f]{encoding = *, family = * }

\newcolumntype{X}[1]{>{\centering\arraybackslash}m{#1}}

\begin{document}

\begin{frontmatter}



\title{Grey Wolf-Based Task Scheduling in Vehicular Fog Computing Systems}


\author[DLS]{Maryam Taghizadeh}
\ead{taghizadehmail@gmail.com \rm(M. Taghizadeh)}
\author[DLS,CorAuth]{Mahmood Ahmadi}
\ead{m.ahmadi@razi.ac.ir \rm(M. Ahmadi)}
\address[DLS]{Department of Computer Engineering and Information Technology, Razi University, Iran.}
\corauth[CorAuth]{Mahmood Ahmadi.}

  
\begin{abstract}
Vehicular fog computing (VFC) can be considered as an important alternative to address the existing challenges in intelligent transportation systems (ITS). The main purpose of VFC is to perform computational tasks through various vehicles. At present, VFCs include powerful computing resources that bring the computational resources nearer to the requesting devices. This paper presents a new algorithm based on meta-heuristic optimization method for task scheduling problem in VFC. The task scheduling in VFC is formulated as a multi-objective optimization problem, which aims to reduce makespan and monetary cost. The proposed method utilizes the grey wolf optimization (GWO) and assigns the different priorities to static and dynamic fog nodes. Dynamic fog nodes represent the parked or moving vehicles and static fog nodes show the stationary servers. Afterwards, the tasks that require the most processing resources are chosen and allocated to fog nodes. The GWO-based method is extensively evaluated in more details. Furthermore, the effectiveness of various parameters in GWO algorithm is analyzed. We also assess the proposed algorithm on real application and random data. The outcomes of our experiments confirm that, in comparison to previous works, our algorithm is capable of offering the lowest monetary cost.

\end{abstract}




\end{frontmatter}


\section{Introduction}

Nowadays, intelligent transportation system (ITS) is an advanced technology resulting many applications \cite{tong2019artificial,giang2016developing}. Meanwhile, connected autonomous vehicles are play a significant role in various ITSs. On the other hand, it is important noting that 5G networks provide imperatively various services in diverse applications \cite{hakak2023autonomous}. As vehicular technologies are advancing, the number of vehicles are increasing in a city.

\textcolor{black}{The demand of data communication, computational capabilities, and storage is increasingly observed. For example, automatic driving generates 2 GB per second of raw sensor data and requires high computation \cite{huang2018parked}. To this end, the use of cloud computing was suggested \cite{ashok2018vehicular}.  However, the rapid growth of data, the increasing demand for real-time processing in smart applications, and quality-of-service deterioration have led to limitations. To address the limitations of cloud infrastructure, fog computing is presented \cite{hu2017survey}. Fog computing brings the computing resources nearer to the requesting devices. Nevertheless, only restricted devices are capable to receive a real-time response due to installation costs fog. Therefore, the enormous cost of installing fog can be reduced by using the computational power of the available vehicles. In addition, it is estimated that in the year 2025, about $\$210-740$ billion/year of potential value will be produced by Internet of Vehicles (IoV) \cite{keshari2022survey}, too. It imposes an extensively burden computation in the data centers while they also cannot meet delay-sensitive applications. Hence, using vehicular networks as effective processing resources is an appealing idea. The use of fog computing in vehicular network emerges the new computing model known as vehicular fog computing \cite{keshari2022survey}.} \textcolor{black}{The VFC addresses the challenges such as reducing latency, offloading data centers, and real-time analytics. They offer numerous benefits, including improved safety, enhanced efficiency, personalized services, smart cities, and autonomous vehicles. To handle demanding tasks like real-time image processing, senor data fusion, and machine learning algorithms, high performance processors are needed. According to IEEE ICC 2018, each self-driving car would be performed instructions powerfully in the forthcoming future, i.e., ten times of computation capability as compared to currently working laptops \cite{sun2018task}.} Therefore, vehicular computing fogs (VFCs) can provide an inexpensive approach for accomplishing diverse user's requests. And more importantly, increasing of the number of vehicles and their power computation can be effective in implementing intelligent cities. By leveraging the computational power of vehicles, the VFC can enable new applications, improve efficiency, and enhance user experience.

The vehicles provide powerful computation resources, while vehicles remain parked approximate 96\% of the time \cite{RAC}. As a result, parked vehicles can be employed as fog nodes in vehicular fog computing and can perform various tasks. Furthermore, these vehicles are able to benefit of their resources. \textcolor{black}{Meantime, vehicular owners are entitled to receive variety of attractive incentives such as free parking, shopping voucher, or free WI-FI in return for lending the computing resources of vehicles to the fog service provider \cite{sookhak2017fog}. For this aim, some incentive mechanisms have been presented \cite{li2019contract,liu2017incentive}.}

\textcolor{black}{In summary, VFCs are enabling a new wave of applications that leverage the computational power and connectivity of vehicles. These applications exhibit several distinct characteristics: real-time processing, context-aware processing, collaborative processing, data-driven decision-making, security, and privacy. These applications are poised to revolutionize various industries, from transportation and logistics to smart cities and autonomous vehicles.}     

The VFC can be a promising computing paradigm because of the following reasons:
\begin{itemize}
\item The presence of a vast and affordable computing capability within vehicles allows for its utilization by other vehicles or computing tasks of users. By 2023, it is estimated that there will be approximately 1.475 billion vehicles worldwide. Assuming an average computing capacity of 10 GFlops per car, the collective computing power within these vehicles amounts to around 14,750 Peta Flops. This impressive figure is 12 times faster than the performance of the most powerful supercomputer. Currently, the Frontier HPE Cray EX235a holds the title of the most powerful supercomputer, boasting 1195 PetaFlops and consuming 22.7 Mwatts of power \cite{ref-new3}. 

\item In comparison to traditional vehicle systems, the progress made in computing technology for vehicle systems has been remarkable. For instance, the ARM Cortex-A5 achieves a performance of 284 MFlops per core, whereas the ARM Cortex-A15 delivers a performance of 1502 MFlops per core, resulting in an overall performance of 33GFlops \cite{ref-new1}. Additionally, the latest ARM Cortex A-78 exhibits an impressive performance of 15.5 GFlops \cite{ref-new2}.

\item The analysis shows that the 95\% of the time the cars are spent in the parking. Therefore, their CPU can be used fully for compute-intensive tasks. In addition, in some places e.g. airport, public parking places. In the Manchester airport 22000 parking places are used.
\item The appearance of the 6G, 5G, and LTE technologies that improve the quality of service metrics and increase the bandwidth requirement to implement vehicle-based computing. In 5G, the goal is to achieve the end-to-end delay of 1 msec, peak data rate of 20 Gbsec, and average data rate of 200 Mbsec. Therefore, the processing of the application with delay-sensitive and bandwidth limitations using vehicle-based computing is overcome. 

\end{itemize}

One of the significant challenges in VFC-based models is resource management and task scheduling for users' requests. The main purpose is to provide a low delay time and the processing cost is appropriate from the user's point of view. Although there are plenty of studies the task scheduling problem in cloud-fog computing environments, few studies have been presented in VFC systems. This paper explores the task scheduling problem in VFC environments. We propose a new algorithm based on a meta-heuristic algorithm to assign resources to tasks. These algorithms are usually suitable to solve this problem due to their global search ability and have also attracted the attention of many researchers \cite{liu2019ant,abd2021advanced,abd2019task}. In this study, we present a method based on the Grey Wolf optimization algorithm to address the task scheduling problem. First, a GWO-based model is presented to assign appropriate resources to different tasks with aiming decrease cost. Second, task scheduling is performed. Experimental results verify the reduction of cost compared to other methods.   
   
In short, the main contributions of this paper are as follows:

\begin{enumerate}
	\item A new approach is proposed for task scheduling in a hybrid cloud-fog environment for vehicular fogs. The proposed approach relies on Grey Wolf optimization algorithm to address resources' assignment to tasks with the aim to decrease makespan and cost. Moreover, task scheduling process is performed based on selecting tasks with more requesting resources, and fog nodes are also labeled by different priorities.   
	\item The proposed algorithm is evaluated for a different number of tasks with various random characteristics. In addition, one real application is assessed for further analysis.  Additionally, modifying the number of static and dynamic fog nodes is examined in the proposed method. We investigate the effectiveness of different parameters of the GWO algorithm in our method.   
\end{enumerate}

The structure of this paper is as follows: Section 2 provides an overview of the previous works. Section 3 explains the basic model and architecture of the proposed system. The proposed algorithm for task scheduling is presented in Section 4. All simulation parameters and the preliminary obtained results are shown in Section 5. Finally, the paper is concluded.

\section{Related work}
In this section, we review recent studies on task scheduling problems for cloud and fog systems. In general, task scheduling is described as how to schedule and allocate various tasks to different virtual machines in a cloud-fog environment. In cloud-fog computing-based systems, task scheduling can be addressed using meta-heuristic \cite{abd2021advanced, saif2023multi} and machine learning algorithms \cite{vemireddy2021fuzzy,kumar2023autonomic}. Although, machine learning-based methods are effective for dynamic environments and improve efficiency, they impose huge complexity and computational cost. Meta-heuristic algorithms can explore a large space and provide the optimal solution appropriately. Therefore, this paper presents an effective method based on meta-heuristic algorithm. In the following, we demonstrate some previous works in task scheduling based on machine learning and meta-heuristic.

\subsection{Machine learning-based methods}
Jamil et al. proposed an intelligent, priority- and deadline-aware resource allocation and task scheduling algorithm named IRATS. The method employed a proximal policy optimization technique and deep reinforcement learning (DRL) to enhance task completion rates and minimize wait time and delay by taking into account task priority and the duration of vehicle links \cite{jamil2023irats}. Ranjan and Sharma proposed a scheduler that utilizes nonlinear mathematical programming for cloud and fog computing. They applied a wavefront cellular learning automata algorithm improved by a genetic algorithm, decreasing energy consumption \cite{jassbi2023improvement}. A Deep Learning Algorithm for Big Data Task Scheduling System (DLA-BDTSS) for Internet of Things and cloud computing applications is presented in another study by Pal et al. \cite{pal2023intelligent}. This system makes use of deep learning to analyze and process a variety of tasks. In addition, Li et al. introduced a Vehicular Task Scheduling Policy Optimization (VTSPO) algorithm that leverages policy-based deep reinforcement learning (DRL) to optimize the performance of vehicular edge computing networks \cite{li2022dependency}.

\subsection{Meta-heuristic-based methods}
Xu et al. \cite{xu2019method} proposed a priority-based task scheduling method using laxity and ant colony in a cloud-fog environment. They attempt to reduce energy consumption and failure rate for scheduling dependent tasks with mixed deadlines. Experiments showed better results compared to GfE, HEFT, and DEACO algorithms. An improved moth search algorithm (MSA) using differential evolution (DE), named MSDE was introduced for task scheduling in cloud computing \cite{abd2019task}. In \cite{nguyen2019evolutionary}, a genetic algorithm was utilized in the cloud-fog system to reduce makespan and monetary cost. 

In another work \cite{movahedi2021efficient}, Mohadi et al. introduced a different architecture for task scheduling. They formulated the optimization problem by integer linear programming (ILP) to reduce energy consumption and time. Then, the problem is addressed using an opposition-based chaotic whale optimization (OppoCWOA) approach. The results have shown better results than the conventional whale optimization method. A bio-inspired hybrid algorithm (NBIHA) was proposed in cloud-fog environments \cite{rafique2019novel}. In this model, a modified particle swarm optimization (MPSO) and a modified cat swarm optimization (MCSO) are combined. Fog nodes are selected through the requested memory and CPU time for each task. The task scheduling in cyber-physical system applications was solved by the TS-MFO method while fog nodes are allocated to each task using the moth flame optimization \cite{ghobaei2020efficient}. The model meets the quality of service (QoS) requirements, and decreases the execution time. 

The research presented in \cite{saif2023multi} has considered a multi-objective grey wolf optimization algorithm to schedule tasks in a fog system as the main aim is to reduce delay and energy. A hybrid method including the fireworks algorithm (FWA) as a meta-heuristic algorithm and HEFT heterogeneous as a heuristic algorithm, called BH-FWA, was presented \cite{yadav2022bi}. To reduce the cost and makespan and to improve the throughput, the authors follow a bi-objective function for scheduling. A summary of the reviewed methods with their details is shown in Table.~\ref{tab:t1}.

\textcolor{black}{In this Table, the main aim is only to show some of the previous works. It can be inferred that there are fewer works in VFC such as \cite{jamil2023irats,jassbi2023improvement,li2022dependency} than systems without VFC. Compared with genetic algorithm (GA) \cite{jassbi2023improvement}, the GWO algorithm requires fewer parameters to tune, making it easier to implement and less prone to overfitting. Additionally, GWO often converges more quickly than GA, leading to faster task scheduling decisions. On the other hand, GWO strikes a good balance between exploration and exploitation, ensuring that it can find both global and local optima. Therefore, this balance allows GWO to explore a wider range of potential solutions, potentially leading to more efficient task allocations. GWO generally requires less computational effort than GA, especially for large-scale problems. In other words, this lower computational cost can translate to reduced energy consumption, which is crucial for battery-powered devices in vehicular fog computing. By effectively balancing exploration and exploitation, GWO can allocate tasks to the most suitable fog nodes, minimizing delays. We can anticipate that GWO is well-suited for handling large-scale task scheduling problems, making it suitable for vehicular fog computing environments with many vehicles and tasks.}

Our method addresses task scheduling in the form of the optimization problem using grey wolf optimization which simply formulates multiple objectives. In sum, various meta-heuristic methods, including genetic algorithm \cite{khiat2024genetic}, ant colony optimization  \cite{sharma2022ant}, and particle swarm \cite{jena2015multi}, have been employed to tackle the scheduling problem. However, meta-heuristic algorithms exhibit several search processes, issues with randomness, limited ability to perform global searches, and a decrease in convergence during later iterations. These factors result in the algorithms being more likely to find local optimum solutions \cite{deng2020energy} and a lack of balance between global and local search \cite{paknejad2021chaotic}. Subsequently, the researchers turn their attention to the Grey wolf optimizer (GWO) because of its notable importance in surpassing the majority of meta-heuristic algorithms. The GWO algorithm relies just on the position of a single vector, resulting in a lower memory requirement as compared to the Particle Swarm Optimization (PSO) method. In contrast to the PSO method, which selects a single best solution from all particles, GWO utilizes the three most optimal solutions to prevent convergence to a local optimum \cite{mirjalili2014grey}. In addition, the GWO algorithm has fewer parameters compared to GA algorithms, resulting in reduced complexity and decreased computing time and energy consumption \cite{mirjalili2014grey}. This paper aims to minimize makespan and monetary cost in vehicular fog computing consisting of parked and mobility vehicles. \textcolor{black}{In VFC, cars can be employed in various modes, including parked, moving, and on/off modes. Cars parked in designated areas can serve as stationary computing nodes. Cars moving on the road can act as mobile edge computing nodes, providing computational resources and connectivity to nearby devices or applications. The moving mode is well-suited for real-time applications that require low latency, such as autonomous driving, traffic management, and emergency response. In on/off mode, cars can optimize their power consumption by turning off or reducing the computational resources when not in use. Also, they can be equipped with mechanisms to respond to specific events or requests. Processors can overheat in extreme temperatures, affecting their performance and reliability. Therefore, effective thermal management systems are crucial to maintain optimal operating conditions.} Here, the parked vehicles are used as fog nodes to contribute the idle computing resources. It is important to mention that vehicles are able to acquire more revenues using sharing their computing resources when they are parked for a long time in different places. \textcolor{black}{A task running can be stopped because of failing a dynamic fog node or moving a vehicular, and leaving the area in VFC. There are approaches to resolving this problem. For example, tasks can be migrated to cloud data centers, and offloaded to other vehicles in the same area \cite{mishra2023collaborative}. Briefly, these factors can be affected by delay, cost, and other metrics. This paper employs parked cars in different areas and } future work, we concentrate on using moving vehicles in task scheduling.

\begin{table*}[h!]
	\caption{Summary of task scheduling based on meta-heuristic and deep learning algorithms}
	\begin{center}
		\resizebox{0.99\textwidth}{!}{  
			\begin{tabular}{|c|c|c|c|c|}
				\hline
				{\textbf{Method}}&\textbf{Optimization algorithm}&\textbf{Metrics}&\textbf{Application}&\textbf{Cloud/fog}\\
				\hline
				\multirow{2}{*}{}&\multirow{3}{*}{Laxity and ant colony system}&\multirow{3}{*}{Energy consumption, Task scheduling failure} &&\\
				\cite{xu2019method}&&&IoT&Cloud-fog\\
				&& rate with mixed deadlines&&\\
				\hline
				&\multirow{3}{*}{Moth search algorithm (MSA), }&&&\\
				MSDE \cite{abd2019task}&&Makespan&Benchmark functions&Cloud\\
				&Differential Evolution&& CEC2005&\\
				\hline
				&\multirow{2}{*}{}&&&\\
				TS-MFO \cite{ghobaei2020efficient}&Moth-flame &Execution time&Cyber-physical system&Fog\\
				&&&&\\
				\hline
				&\multirow{2}{*}&&&\\
				NBIHA \cite{rafique2019novel}&Modified particle swarm,&Energy consumption, execution&IoT&Cloud-fog\\
				&modified cat swarm &time, average response time&&\\
				\hline
				\multirow{2}{*}&&&&\\
				OppoCWOA &Opposition-based chaotic&Time, energy consumption&IoT&Fog\\
				\cite{movahedi2021efficient}&whale optimization&&&\\
				\hline
				&&&&\\
				\cite{nguyen2019evolutionary}&{Genetic algorithm}&Execution time, monetary cost&IoT&Cloud-fog\\
				&&&&\\
				\hline
				&&&&\\
				MGWO \cite{saif2023multi}&Multi-objectives grey wolf &Delay, energy consumption&IoT&Cloud-fog\\
				&optimization&&&\\
				\hline
				\multirow{2}{*}&&&&\\
				BH-FWA \cite{yadav2022bi} &{Fireworks algorithm (FWA),}&Makespan, cost,&IoT&fog\\
				& Heterogeneous earliest finish time&throughput&&\\
				\hline
				\multirow{2}{*}&\multirow{2}{*}&&&\\
				IRATS \cite{jamil2023irats}&proximal policy optimization,&Task completion, wait time,&Vehicular&fog\\
				&deep reinforcement learning (DRL) &delay&&\\

				\hline
				\multirow{2}{*}&\multirow{2}{*}&&&\\
				\cite{jassbi2023improvement}&Wavefront cellular learning automation,&Energy consumption&Vehicular&Cloud-fog\\
				&genetic algorithm &&&\\
				
				\hline
				\multirow{2}{*}&\multirow{2}{*}&&&\\
				DLA-BOTSS&Deep learning algorithm&Energy cost, makespan,&IoT&Cloud\\
				\cite{pal2023intelligent}& &resource consumption&&\\
				\hline
				\multirow{2}{*}&\multirow{2}{*}&&&\\
				VTSPO \cite{li2022dependency}&DRL&Performance&Vehicular&Fog\\
				& &&&\\
				\hline
				
		\end{tabular}}
		\label{tab:t1}
	\end{center}
\end{table*}

\section{System model}
\label{S3}
This section demonstrates the system architecture in detail as illustrated in Figure.~\ref{system}, followed by the problem formulation. Table.~\ref{tab:t2} contains a collection of the notifications and symbols used for the convenience of readers.  

\begin{table}[h!]
	\caption{Defined notations }
	\begin{center}\hspace*{-1cm}
		\scalebox{0.70}{
			\begin{tabular}{|c|c|}
				\hline
				\textbf{Notation}&\textbf{Description}\\
				\hline
				N&Task number\\
				\hline
				S&Static fog node number\\
				\hline
				D&Dynamic fog node number\\
				\hline
				C&Cloud server number\\
				\hline
				$PT_{i}$&Processing time of $T_{i}$\\
				\hline
				$ET_{i}$&Completion time $T_{i}$\\
				\hline
				$CT$&Completion time (Makespan)\\
				\hline
				$CS_{i}$&Static processing core number allocated to $T_{i}$\\
				\hline
				$CD_{i}$&Dynamic processing core number allocated to $T_{i}$\\
				\hline
				$CC_{i}$&Cloud processing core number allocated to $T_{i}$\\
				\hline
				$FS$&Monetary cost of each static fog node in (currency/second) \\
				\hline
				$FD$&Monetary cost in each dynamic fog node in (currency/second)\\
				\hline
				$FC$&Monetary cost in cloud servers  in (currency/second)\\
				\hline
				$CSD_{i}$&Monetary cost $T_{i}$ in dynamic and static fog node  in (currency/second)\\
				\hline
				$CCld_{i}$&Monetary cost $T_{i}$ in cloud servers  in (currency/second)\\
				\hline
				$CostT_{i}$&Transfer monetary cost  to cloud server in (currency/second)\\
				\hline
				$Cost$&Total monetary cost for all tasks (currency)\\
				\hline
		\end{tabular}}
		\label{tab:t2}
	\end{center}
\end{table}
 
\subsection{System architecture}
\begin{figure}[h!]
	\begin{center}
		\includegraphics[width=1\linewidth,height=0.65\columnwidth]{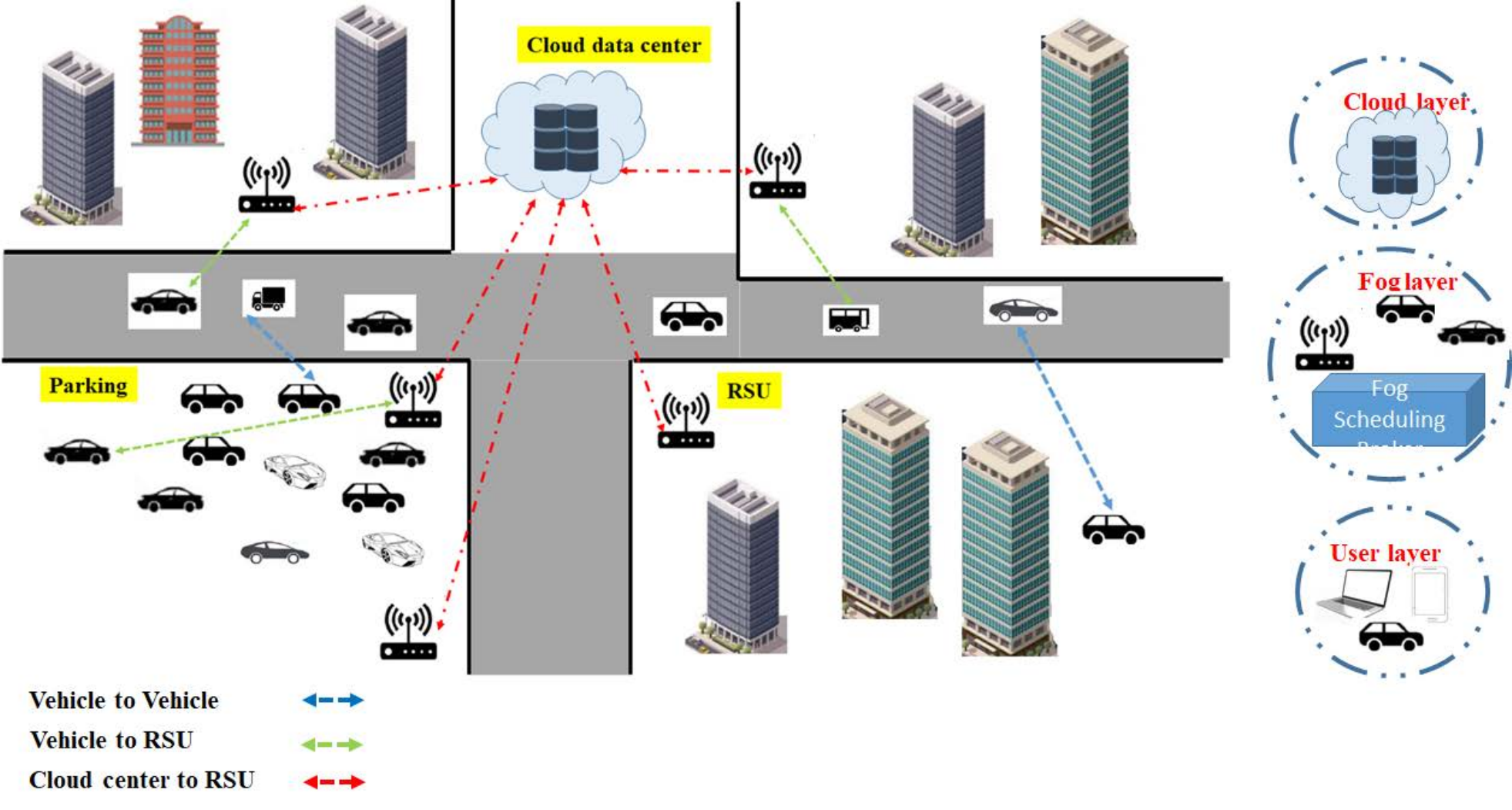}
		\caption{System model}
		\label{system}
	\end{center}
\end{figure}

As shown in Figure.~\ref{system}, the proposed network architecture includes the fog layer, cloud layer, and user layer. Further explanations of each layer are provided below. At the user layer, wireless channels are used for communication between end users. These users have direct interaction with the fog layer. The broker component, which is integrated into the fog layer, receives the data produced by users and employs it to schedule tasks and assign resources. It chooses fog nodes or cloud servers based on the proposed method that is given to carry out computations.

\textbf{Fog layer:} The fog layer comprises a collection of fog nodes, which are heterogeneous computing devices including switches, routers, static servers, and dynamic servers. Each fog node is equipped with one or several virtual machines, each with the capability to execute tasks. In this paper, we employ vehicles that are either in motion or stationary as fog nodes. It is postulated that in the current layer, there exists a broker that takes requests from the user layer and assumes the responsibility of scheduling tasks and distributing resources. This unit collects data on the allocation of processing resources, computing power, and processing expenses of virtual machines. Utilizing the provided data and the suggested algorithm, it conducts resource allocation and task scheduling. Hence, it allocates jobs among virtual machines. Upon completion of the assigned tasks by several virtual machines, the outcomes are conveyed to the users.

It is worth mentioning that there are Roadside Units (RSUs) strategically placed along the roadways to facilitate network access. The unit in vehicular fog computing systems can be displayed in two distinct modes: decoupling mode and coupling mode. In the first scenario, there exists an intermediary layer where the roadside units are distinct and isolated from the fog machines. In the second mode, a combination of roadside units and fog machines are situated together \cite{yu2018deployment}. This paper assumes the utilization of the coupling mode. Consequently, the architecture no longer has a distinct layer known as the RSU layer.

\textbf{Cloud layer:} The layer comprises a collection of virtual machines that possess superior processing and computing capabilities compared to fog layer computing systems. This sector includes cloud data centers.

\textbf{User layer:} In this layer, a range of devices, including mobile phones and various sensors, are outfitted with diverse communication technologies. \textcolor{black}{Different types of data are derived from cameras, radar, LiDAR, GPS, and other sensors on the vehicle. It also includes data on traffic flow and congestion that can be utilized to optimize traffic management, improve transportation efficiency, and reduce congestion. The user data can consist of user preferences, behavior, and interactions with the vehicle. Lastly,} cloud computing system must carry out the derived requests, and the consumers then receive the results.

The objective is to allocate tasks to the computing nodes in the fog and cloud layers to minimize the completion time of the final task and the monetary cost.

\subsection{Problem formulation}
Let's assume that the total amount of static resources we utilize as edge machines is denoted as $S$. Multiple virtual machines are housed within each processing node. To simplify, we assume that the features of each edge machine are identical. Let's define the number of resources that are in motion, including cars, as $D$. Each of these resources can contain numerous virtual machines or computing resources.

Every task necessitates an exact number of computing resources for its completion. The time and resource requirements for each task are stated to facilitate work. Tasks are independent and can be concurrently executed. Furthermore, it is possible to allocate each task to distinct computational resources. Indeed, each task has the capability to execute concurrently on separate machines. The quantity of tasks is denoted as $N$ and represented as $\{T_{0},T_{1},...,T_{N-1}\}$. Each task $T_{i}$ possesses specific attributes, including the number of computational functions needed and the time required to access the resource. These attributes are specified as a 3-tuple $(i,R_{i},te_{i})$, where the value \textit{i} represents the task index and $R_{i}$ represents the computing resources required by task $i$. The duration of resource ownership is equivalent to $te$.

\textbf{End time of the last task:}
The task completion time is equivalent to the duration required to finish the task. This duration encompasses both wait and processing times. The duration required to finish a task is determined in the following manner:

\begin{equation}
ET_{i}=PT_{i}+WT_{i}
\label{eq1}
\end{equation}

The execution time of task \textit{i} is equivalent to the sum amount of its processing time and wait time. It is presumed that the task is initially verified and performed on fog nodes. In the case that the fog nodes are incapable of executing, the task is sent to the cloud layer. We disregard the transit time between the static and dynamic fog nodes because it is convenient to assume it is little. The processing time for task \textit{i} is determined using the following method:

\begin{equation}
PT_{i}=te_{i}\times max(X_{ij})
\label{eq2}
\end{equation}

The $X$ matrix contains either zero or one term. The value of $X_{ij}$ in the matrix will be one if the fog node $j$ (whether it's static or dynamic) is assigned to a task $i$, and zero otherwise.

\begin{equation}
X_{ij}=\begin{cases}
1  \hspace{3pt} if \hspace{3pt} V_{j}  \hspace{3pt} is \hspace{3pt}  assigned \hspace{3pt} to \hspace{3pt} T_{i} \\
0   \hspace{3pt} else \\
\end{cases}
\label{eq3}
\end{equation}
The transfer time to the cloud increases when a task is delegated to the cloud layer for execution; this can be determined individually and incorporated into equation (2). The overall completion time (makespan) is now equivalent to the completion time of the final task, assuming that all tasks were arranged for execution at the start of the time interval as written in Eq.~\ref{eq4}.

\begin{equation}
CT=max(ET_{i}) , i=\{1,2,...,N\}
\label{eq4}
\end{equation}

As the time value decreases, the system's efficiency increases, resulting in improved service quality and satisfaction for users. 

\textbf{Cost:} The expenses of executing a task include numerous services that are offered to the user for the execution of the task over some time, and this amount of cost is decided depending on the duration of taking over the computer systems. It is postulated that this expense is proportional to the quantity of computing units designated for execution. The equation below illustrates the cost associated with executing a task on both static and dynamic nodes:

\begin{equation}
\begin{aligned}
CSD_{i}=\sum_{j=1}^{S} CS_{j} \times PT_{ij} \times FS \times X_{ij} +\\ 
\sum_{j=1}^{D} CD_{j} \times PT_{ij} \times FD \times X_{ij}
\label{eq5}
\end{aligned}
\end{equation}

Equation.~\ref{eq5} estimates the total cost for fog nodes by multiplying the number of processing cores and the requested duration time for each task with the default cost per fog node. For dynamic fog nodes, the amount of cost is likewise obtained in this way. Then, for all static and dynamic fog nodes, these values are calculated, and their sum is taken as the cost of the static and dynamic parts of fog computing.

If a task or a portion of it is performed on the cloud layer and cloud data centers, the total cost includes both the cost of computing on the cloud and the cost of sending data to the cloud.

\begin{equation}
CCld_{i}= \sum_{j=1}^{C} CC_{j} \times PT_{ij} \times FC + CostT_{ij}
\label{eq6}
\end{equation}

\textit{CS}, \textit{CD}, and \textit{CC} are all positive integers. The final monetary cost for executing all tasks is the sum of Eq.~\ref{eq5} and Eq.~\ref{eq6} as acquired by:

\begin{equation}
Cost= \sum_{i=1}^{N} CSD_{i} +CCld_{i}
\label{eq7}
\end{equation}

\textbf{Objective function:} The objective of the optimization challenge is to reduce both the completion time of the final task (Eq.~\ref{eq4}) and the monetary cost of the system (Eq.~\ref{eq7}). We convert the multi-objective optimization problem into a single-objective optimization problem by the utilization of the weighted sum approach, and subsequently solve it.

\begin{equation}
F=\alpha \times Cost + \beta \times CT
\label{eq8}
\end{equation}

We seek to minimize the function \textit{F}. The following conditions apply:
\begin{equation}
\alpha +\beta=1
\label{eq9}
\end{equation}

The point to be noted here is the selection of suitable values for parameters $\alpha$ and $\beta$. The selected value must be in the range (1, 0). These two parameters can be quantified by experimental or learning-based methods. Different conditions for $\alpha$ and $\beta$ including (0.25, 0.75), (0.75, 0.25), and (0.5, 0.5) can be considered respectively. For ease of work in the first implementation, we consider the equal value for $\alpha$ and $\beta$ in Eq.~\ref{eq8}.

Considering that the two values in the objective function are not of the same type, we normalize them and replace them in Eq.~\ref{eq8} \cite{nguyen2019evolutionary}. For the value of the monetary cost, the optimal value is when the tasks are executed on static servers or dynamic fog nodes, and there is no need to execute them on the cloud, and therefore the transfer cost will not be imposed on the system. The following equation shows the optimal amount of monetary cost.

\begin{equation}
Cost_{M}=\sum_{i=1}^{N} R_{i} \times te_{i} \times min(FS, FD)
\label{eq10}
\end{equation}

For the termination time of the last task, the minimum amount of time is obtained when the wait time is equal to zero and all tasks are executed on cloud computing nodes and the cost of transferring to the cloud is not needed.

\begin{equation}
CT_{M}=minET_{i}
\label{eq11}
\end{equation}
So the objective function will be like this and we are looking to maximize it.
\begin{equation}
F= \alpha \times Cost_{M} / Cost + \beta \times CT_{M}/CT
\label{eq12}
\end{equation}

The following restrictions are also in place:
\begin{equation}
\sum_{j=1}^{S+D+C} X_{ij} \ge 1
\label{eq13}
\end{equation}

\begin{equation}
0 \le \sum_{i=1}^{N} X_{ij} \le N 	
\label{eq14}
\end{equation}
The proviso stated in Eq.~\ref{eq13} shows that each task is assigned computing resources that can execute its commands. Eq.~\ref{eq14} states that the total number of tasks assigned to a computing node j will be at most N (equal to the number of entered tasks).

\section{Proposed method}
\label{s4}
This section demonstrates the proposed method in more details. We have given priority to static fog nodes, dynamic fog nodes and then computing machines in the cloud layer, respectively. \textcolor{black}{Generally, tasks can be allocated to cars based on their proximity to the data source, computational capabilities, network connectivity, and battery level. Tasks can be priorities based on their importance, urgency, and resource requirements. Herein, the scheduling algorithm attempts to assign tasks into virtual machines in dynamic and static fog nodes while the GWO-based algorithm has obtained them according to initialized parameters in the GWO algorithm and using maximizing objective function as written in Eq.~\ref{eq12}. Now, provided that a vehicular is considered for assigning more the one task concurrently, the algorithm gives priority to tasks with more computation requirements.} We use the grey wolf meta-initiative algorithm to allocate resources to tasks. We solve the objective function proposed in the previous section using this algorithm. First, the grey wolf optimizer is explained, next, we describe the proposed algorithm for task scheduling.

\subsection{Grey wolf optimization (GWO) algorithm}    
\begin{figure}[h]
	\begin{center}
		\includegraphics[scale=0.8]{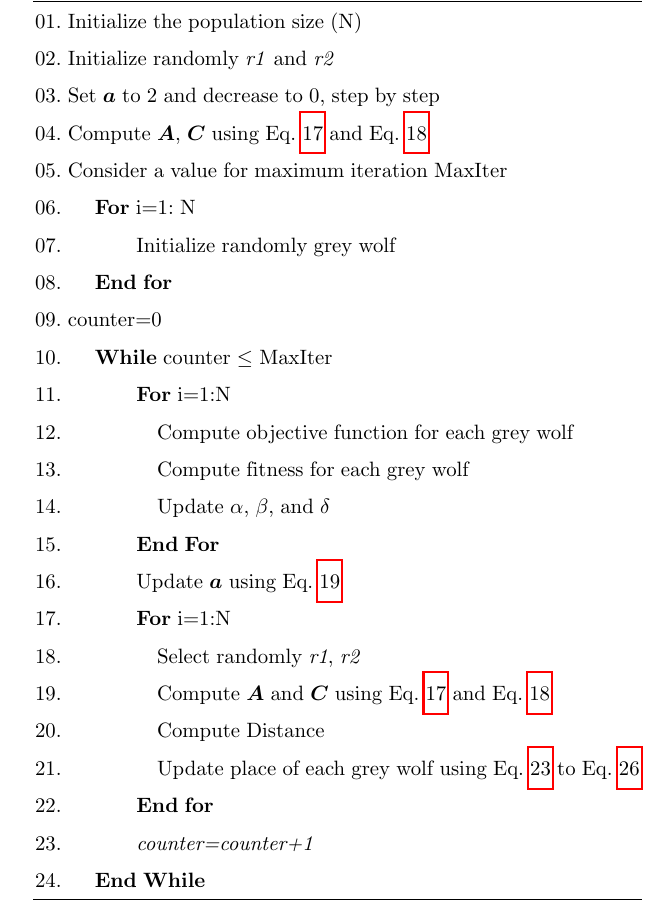}
		\caption{Pseudo code of grey wolf optimization (GWO)}
		\label{algo1}
	\end{center}
\end{figure}

The grey wolf algorithm is a meta-heuristic algorithm that takes inspiration from the hierarchical structure and social behavior of grey wolves during hunting \cite{mirjalili2014grey}. The method is characterized by its population-based nature and its straightforward procedure, which allows for easy scalability to tackle large-scale challenges. Within each group, the members are divided into four distinct categories: alpha wolves (who serve as the leaders of the group), beta wolves (who assist the alpha wolves in making decisions), delta wolves (comprising older wolves, hunters, and wolves responsible for caring for the young), and omega wolves (who occupy the lowest position in the hierarchical structure and have no involvement in decision-making). The algorithm consists of 3 steps and its pseudo-code is shown in Figure \ref{algo1}.

The initial phase entails the act of seeing the prey, subsequently tracking it, and ultimately engaging in pursuit. In the second phase, the predator engages in activities such as approaching the prey, encircling it, and diverting its attention until it ends its movement. The third phase covers the act of hunting.

The process of optimization is achieved by utilizing beta, alpha, and delta wolves. The algorithm is primarily driven by a hypothetical alpha wolf, with additional contributions from subordinate beta and delta wolves. The remaining wolves are regarded as their supporters. Below is the mathematical representation of the algorithm.

The best solution is $\alpha$, and the second and third solutions are $\beta$, and $\delta$, respectively. The remaining solutions are part of the $\omega$ group. During the initial phase, the hunt is approached and surrounded by the following relations:

\begin{equation}
\vec{D}=|\vec{C} \times \vec{X_{p}(t)}-\vec{X_{t}}|
\label{eq15}
\end{equation}

\begin{equation}
\vec{X}(t+1)=\vec{X_{p}(t)}-\vec{A} \times \vec{D}
\label{eq16}
\end{equation}

where, the index \textit{t} denotes the iterative steps of the method. The variable $\vec{X_{p}}$ represents the location of the hunt, whereas the variable $\vec{X}$ represents the location of the wolf. The vector $\vec{D}$ represents the distance between the wolf and the prey. The vectors $\vec{A}$ and $\vec{C}$ can be written as:

\begin{equation}
\vec{A}=2\vec{a}\times \vec{r1} - \vec{a} 
\label{eq17}
\end{equation}

\begin{equation}
\vec{C}=2 \times \vec{r2}
\label{eq18}
\end{equation}
  
Here, $\vec{a}$ decreases linearly from 2 to zero. The values of \textit{r1} and \textit{r2} are random vectors between zero and one. The $\vec{a}$ is calculated as:

\begin{equation}
\vec{a}=2-2\times (t \div Iteration)
\label{eq19}
\end{equation}  

During the hunting phase, the three most optimal solutions are found. The position of these solutions will be revised as:

\begin{equation}
\vec{D_{\alpha}}=|\vec{C1} \times \vec{X_{\alpha}}-\vec{X}|
\label{eq20}
\end{equation}

\begin{equation}
\vec{D_{\beta}}=|\vec{C1} \times \vec{X_{\beta}}-\vec{X}|
\label{eq21}
\end{equation}

\begin{equation}
\vec{D_{\delta}}=|\vec{C1} \times \vec{X_{\delta}}-\vec{X}|
\label{eq22}
\end{equation}

\begin{equation}
\vec{X1}=\vec{X_{\alpha}}-\vec{A1}\times \vec{D_{\alpha}}
\label{eq23}
\end{equation}

\begin{equation}
\vec{X2}=\vec{X_{\beta}}-\vec{A2}\times \vec{D_{\beta}}
\label{eq24}
\end{equation}

\begin{equation}
\vec{X3}=\vec{X_{\delta}}-\vec{A3}\times \vec{D_{\delta}}
\label{eq25}
\end{equation}

\begin{equation}
\vec{X}(t+1)=(\vec{X1}+\vec{X2}+\vec{X3})/3
\label{eq26}
\end{equation}

Upon iterating the algorithm for the set number of repetitions, the optimal solution will be chosen as the level $\alpha$.

\subsection{Task scheduling algorithm}
\begin{figure*}[h]
	\begin{center}
		\includegraphics[scale=1]{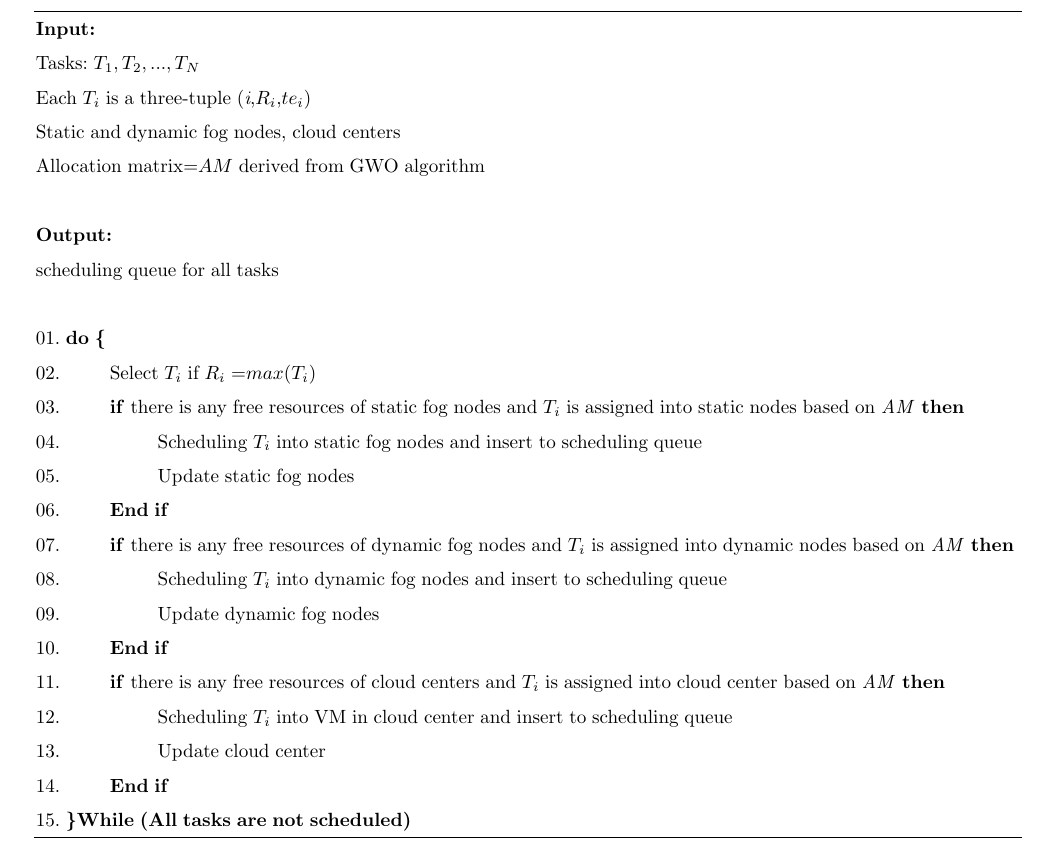}
		\caption{Pseudo code of task scheduling algorithm}
		\label{algo2}
	\end{center}
\end{figure*}

The proposed task scheduling algorithm is presented in details in Figure \ref{algo2}. The algorithm selects a task requesting the maximum number of processing cores. The allocation matrix provided by the GWO algorithm is explored and based on it all tasks are scheduled. This process is repeated till that all tasks are investigated. In sum, the algorithm consists of three parts in a loop structure. First, a task demanding a maximum number of resources is selected (Line 02). Then, the first part examines static fog nodes (Line 03). The second and third parts test dynamic fog nodes and cloud centers, respectively (Lines 07 and 11). This process is continued to all tasks are taken in the scheduling queue (Line 15).

\subsection{Time complexity}
The time complexity of the GWO is based on four phases, including the initialization, the computation of control parameter, the updating of the agent's position, and the evaluation of fitness. In each phase, the complexity can be written as: $O(N\times m)$. Therefore, the sum time complexity is $O(N \times m)$. On the other hand, the algorithm is performed for a maximum iteration $(MaxIter)$. Hence, the total complexity is to equal $O(N\times m \times MaxIter)$.

\section{Experimental results}
\label{s5}
This Section describes experimental results acquired by our implementation and includes four subsections. Subsection ~\ref{s51} demonstrates more details of implementation. Subsection ~\ref{s52} investigates the GWO algorithm for different conditions. \textcolor{black}{The GWO-based algorithm is examined based on its primary parameters. Experimental results show the effectiveness of three parameters, including iteration, a, and search agent. Then, we examined important metrics by considering the best parameters in the GWO-based proposed algorithm, and compared them to previous works in} subsection ~\ref{s53}. \textcolor{black}{Additionally, the paper evaluates the effectiveness of the number of fog nodes on evaluation metrics. Similarly, the effectiveness of coefficients in the cost function is analyzed. To verify our results, the proposed algorithm also carries out the proposed method for a scientific workflow in } subsection ~\ref{s54}.

\subsection{Implementation parameters}
\label{s51}
To implement the proposed method, Python programming \textcolor{black}{and MATLAB software has been used. Tasks are randomly generated in MATLAB, and the grey wolf algorithm is also implemented using MATLAB. To measure all metrics, python programming is utilized. The general system used to implement codes is based on an Intel processor Core i7. To evaluate the proposed model,} two scenarios are supposed to evaluate our method. The first scenario supposes constant values for the number of fog nodes and changes the number of tasks. In the second scenario, we changed the number of fog nodes. In contrast, the same tasks were handed over to the system for scheduling and execution.
The first scenario considers 15 fog nodes and 2 cloud data centers. Fog nodes and cloud data centers are assumed to be heterogeneous with different computing capabilities. Fog nodes include 5 static and 10 dynamic fog nodes. \textcolor{black}{For the number of tasks, different numbers ranging from 10 to 100 tasks have been investigated. We have assumed that each task declares the number of requested processing cores. The characteristics of tasks are chosen randomly similar to \cite{saif2023multi,mao2023demand,yin2018tasks}. In our experiments, we performed all simulations in a geographical area of 9000 $\times$ 9000 $m^{2}$ \cite{sethi2023feddove}.} Also, the required during time for the computing resource is announced in the request. For this reason, we calculate the execution time based on the time demand. Tasks are independent of each other and can be executed in parallel. \textcolor{black}{Notice that the power of the processor in vehicular is important. The arm processors including Arm Cortex-A, Cortex-R, and Cortex-M can be installed in vehicles \cite{bos2022post}. We consider the power of processors based on Arm Cortex-A7 which is 1551 MIPS in our implementation. }In this paper, it is assumed that all tasks are generated and delivered to the scheduler at the beginning of execution. In order to increase the reliability, the experiments were repeated 5 times and their average is reported. Moreover, a real application is examined by the proposed method. Simulation parameters are shown in Table.~\ref{tab:simulation}. In the following, the results are presented.

\begin{table*}[h!]
	\caption{Simulation parameters }
	\begin{center}\hspace*{-1cm}
		\resizebox{0.99\textwidth}{!}{  
			\begin{tabular}{|c|c|}
				\hline
				\textbf{Parameter}&\textbf{Value}\\
				\hline
				\# tasks&[10,20,30,40,50,60,70,80,90,100]\\
				\# cloud center &2\\
				\# static fog node &[5,7]\\
				\# dynamic fog node&[10,12]\\
				\# Computation resources in each VM of static fog nodes&[2, 5, 6, 1, 2]\\
				\# Computation resources in each VM of dynamic fog nodes&[5,2,3,4,2,4,3,5,1,4]\\
				\# Computation resources in each VM of cloud data center&[20,12]\\
				\hline
		\end{tabular}}
		\label{tab:simulation}
	\end{center}
\end{table*}

\subsection{GWO results}
\label{s52}

\begin{figure}[h]
	\begin{center}
		\includegraphics[width=1\linewidth,height=0.85\columnwidth]{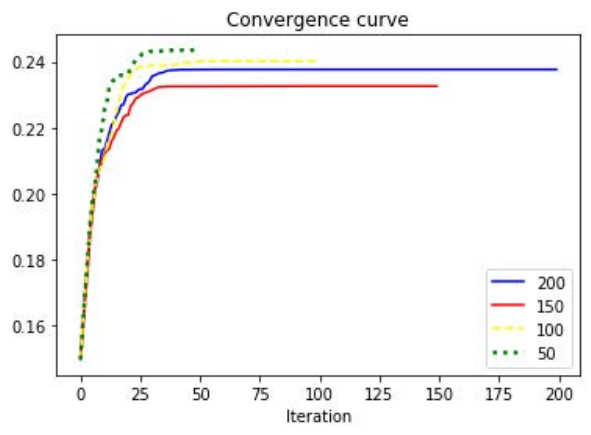}
		\caption{Effect of different values at iteration parameter in GWO algorithm}
		\label{Iteration}
	\end{center}
\end{figure}

\begin{figure}[h]
	\begin{center}
		\includegraphics[width=1\linewidth,height=0.85\columnwidth]{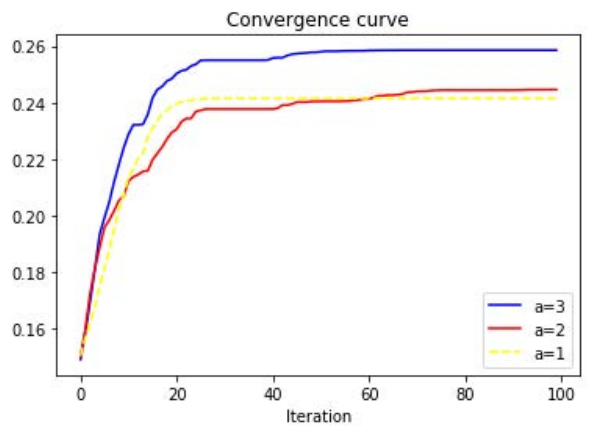}
		\caption{Effect of different values at \textit{a} parameter in GWO algorithm}
		\label{a}
	\end{center}
\end{figure}

\begin{figure}[h]
	\begin{center}
		\includegraphics[width=1\linewidth,height=0.85\columnwidth]{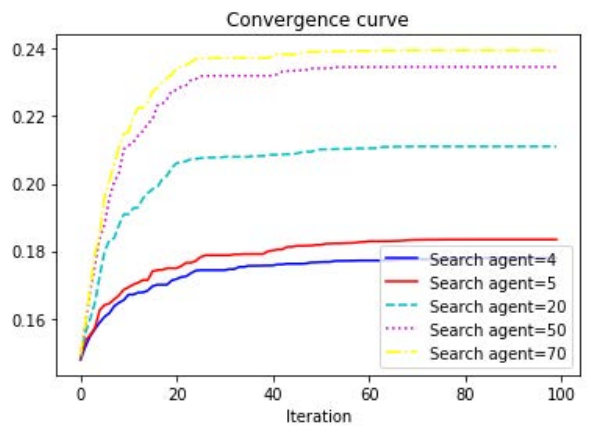}
		\caption{Effect of different values at \textit{Search agent} parameter in GWO algorithm}
		\label{Search}
	\end{center}
\end{figure}

As mentioned earlier, the GWO algorithm has three primary parameters, including \textit{iteration}, \textit{search agent}, and $\vec{a}$ as convergence coefficient. Hence, first, we explore the effect of different parameters in the GWO algorithm on convergence rate. For this purpose, the number of tasks is supposed to be 50 and modify values of GWO parameters, and to implement the proposed method, we use different values for \textit{search agent}, \textit{iteration}, and $\vec{a}$ parameters in GWO algorithm as shown in Table.~\ref{tab:simulationGWO}. \textcolor{black}{According to the literature, a maximum number of iterations are usually between 50-1000, and the number of wolves is selected from [4, 100]. As shown in Figure 3, a parameter is defined in the range [2, 0]. In the same way, the proposed algorithm examines diverse values for all parameters. It should be mentioned that we explored greater value at the convergence coefficient in GWO contrary to the primary version of the GWO algorithm. Then, the convergence curve is drawn based on diverse values for parameters in GWO. Examining the convergence curve resulted in selecting optimal values for the primary parameters of the GWO algorithm. Of course, the performance of the algorithm using all metrics is computed with defined parameter settings.} The convergence curves at different conditions are shown in Figures.~\ref{Iteration}, ~\ref{a}, and ~\ref{Search}. The effect of various \textit{iterations} on the convergence curve is shown in Figure.~\ref{Iteration}. This figure illustrates that the algorithm convergence becomes slower as the number of iterations increases. Also, the scores obtained for fewer repetitions are a larger number. Additionally, Figure.~\ref{a} shows the effect of the $\vec{a}$ parameter in the GWO algorithm. It is clear that the algorithm shows a faster convergence for a smaller value of the $\vec{a}$ parameter. Moreover, the GWO algorithm was simulated using different values for \textit{search agent} parameter as drawn in Figure.~\ref{Search}. As can be observed, by increasing the value of the \textit{search agent} parameter, the obtained score is improved and also the algorithm converges quickly.

\begin{table}[h!]
	\caption{Simulation parameters in GWO algorithm }
	\begin{center}\hspace*{-1cm}
		\scalebox{1.0}{
			\begin{tabular}{|l|c|}
				\hline
				\textbf{Parameter}&\textbf{Value}\\
				\hline
				\textit{Search agent} &[4,5,20,50,70]\\
				\textit{a} &[1,2,3]\\
				\textit{Iteration} &[50,100,150,200]\\
				\hline
		\end{tabular}}
		\label{tab:simulationGWO}
	\end{center}
\end{table}

Second, the paper evaluates different metrics for the proposed algorithm based on considering different values for GWO parameters. We calculate different metrics such as cost function based on Eq.~\ref{eq12}, cost, makespan, and wait time. The results are shown and discussed as below.  

Table.~\ref{tab:res1} shows the values of various metrics for diverse iterations in the GWO algorithm. The main goal is to minimize the cost, wait time, and makespan and maximize the cost function. As shown in the table.~\ref{tab:res1}, iteration values of 50 and 100 show better results for makespan, wait time, cost, and cost function. 

\begin{table}[h!]
	\caption{Different metrics based on various values of \textit{iteration}}
	\begin{center}
		\scalebox{0.85}{
			\begin{tabular}{|c|c|c|c|c|}
				\hline
				\textbf{Iteration }&\textbf{Cost function }&\textbf{Cost}&\textbf{Makespan}&\textbf{Wait }\\
				\hline
				50&0.421071133&44239.2&\textbf{18246.8}&\textbf{17468.8}\\
				\hline
				100&\textbf{0.430391249}&\textbf{43277.2}&20367.2&19653\\
				\hline
				150&0.418652981	&44483.4&19758.8&18911.8\\
				\hline
				200&0.422850081	&44044.8&20309&19463.2\\	
				\hline
		\end{tabular}}
		\label{tab:res1}
	\end{center}
\end{table}

Table.~\ref{tab:res2} shows the measured metrics for different values at $\vec{a}$ parameter.  Again, the experiment was repeated by setting \textit{search agent} to 70. The experimental results show suitable improvement by considering 2 and 1 as $\vec{a}$.

\begin{table}[htb!]
	\caption{Different metrics based on various values of $\vec{a}$}
	\begin{center}
		\scalebox{0.95}{
			\begin{tabular}{|c|c|c|c|c|}
				\hline
				$\vec{a}$&\textbf{Cost function }&\textbf{Cost}&\textbf{Makespan}&\textbf{Wait }\\
				\hline
				3&0.420295556&44292.4&22848.8&21997.6\\
				\hline
				2&\textbf{0.434505}&\textbf{42853}&20920.8&20086.8\\
				\hline
				1&0.42708741&43614.8&\textbf{17182}&\textbf{16469.8}\\
				\hline
				
		\end{tabular}}
		\label{tab:res2}
	\end{center}
\end{table}

The results are provided in Table.~\ref{tab:res3} for different values at \textit{search agent}. As observed from Table.~\ref{tab:res3}, the high value at \textit{search agent} results in improving metrics.

\begin{table}[h!]
	\caption{Different metrics based on various values of \textit{search agent} parameter}
	\begin{center}
		\scalebox{0.85}{
			\begin{tabular}{|c|c|c|c|c|}
				\hline
				\textbf{Search agent }&\textbf{Cost function }&\textbf{Cost}&\textbf{Makespan}&\textbf{Wait }\\
				\hline
				4&0.393055572&47340.2&28330	&27386\\
				\hline
				5&0.401996722&46310.4&26660.6&25737.6\\
				\hline
				20&0.416632225&44702&21473.6&20637.4\\
				\hline
				50&0.422959189&44049.8&20277.2&19406.2\\
				\hline
				70&\textbf{0.425191952}&\textbf{43799.4}&\textbf{20172.6}&\textbf{19278}\\
				
				\hline
		\end{tabular}}
		\label{tab:res3}
	\end{center}
\end{table}

\textcolor{black}{The findings show that the suitable number of iterations is between 50 and 100 and both Figure.~\ref{Iteration} and Table.~\ref{tab:res1} verify it. According to Table.~\ref{tab:res2} and Figure.~\ref{a}, the best selection for convergence coefficient a can be 1 and 2. The optimal parameter settings can vary depending on the specific application and problem characteristics. In the following, the benefit value at search agent is initiated by 70 by examining Table.~\ref{tab:res3} and Figure.~\ref{Search}. }

\textcolor{black}{As stated before, the best parameter settings can depend on the specific application and problem characteristics. However, some general guidelines can be followed as: (1). For large-scale problems, larger population sizes and more iterations can be identified. Careful tuning of the convergence coefficient is essential to balance exploration and exploitation. (2). For small-scale problems, smaller population sizes and fewer iterations are reasonable. Lastly, by carefully considering these factors and experimenting with different parameter settings, it is possible to optimize the performance of GWO for a wide range of applications.}

\subsection{Results and Comparison}
\label{s53}
As mentioned before, this paper explores three metrics as follows: makespan, wait, and cost. All results are compared to random method and heuristic methods. The random method assigns resources to tasks randomly, while the heuristic methods operate based on a set of specific rules. The heuristic algorithms give priority to selecting a task with the maximum (minimum) expected resources to execute by heuristic rule \cite{taghizadeh2024heuristic}. Also, the Minimum Execution Time (MET) algorithm \cite{freund1998scheduling} assigns VM to the task that shows minimum execution time. Here, we consider the MET heuristic algorithm as an algorithm that requests minimum duration time. Additionally, we explore the effects of the number of fog nodes and coefficients in Eq.~\ref{eq12} for the proposed algorithm.

\begin{figure*}[h]
	\begin{center}
		\includegraphics[scale=0.55]{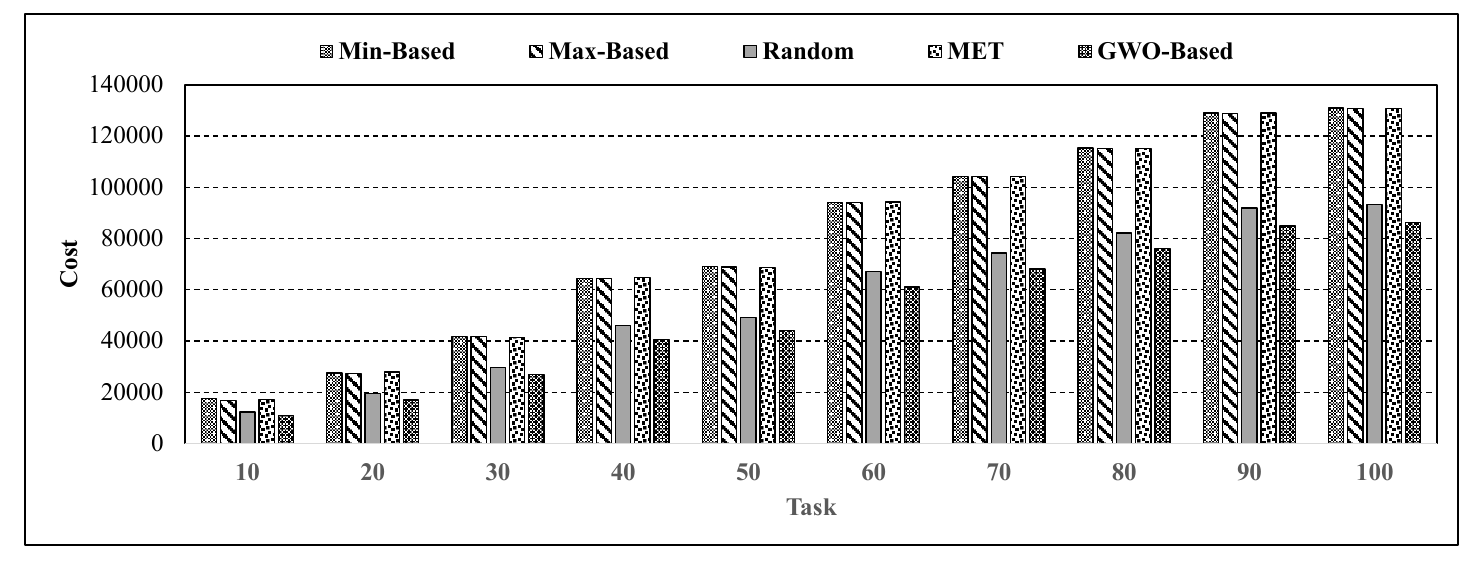}
		\caption{Comparison cost for the proposed method with different methods}
		\label{comparecost}
	\end{center}
\end{figure*}

\textbf{Comparison:} We compare the proposed method with four methods including random, minimum-based, maximum-based \cite{taghizadeh2024heuristic}, and minimum execution time (MET). Results are shown in Figure.~\ref{comparecost}. The results verify the supremacy of the grey wolf-based proposed method in comparison to other methods in term of cost metric. We generally supposed that the cost metric is based on currency. In makespan and wait metrics (Tables.~\ref{tab:resgwo} and ~\ref{tab:waitgwo}), the proposed method is superior to the random method, but the heuristic methods show more suitable time. 

\begin{table}[h!]
	\caption{Makespan metric in different methods }
	\begin{center}
		\scalebox{0.75}{
			\begin{tabular}{|c|c|c|c|c|c|}
				\hline
				\textbf{\# Task }&\textbf{Min-Based }&\textbf{Max-Based}&\textbf{Random}&\textbf{MET }&\textbf{GWO-Based}\\
				\hline
				10&442&	441&7652.4&447&	5970\\
				\hline
				20&623&	654	&11933.4&653&6588\\
				\hline
				30&974	&973&18665&989&13660	\\
				\hline
				40&1479	&1478&29790.4&1462&	18524\\	
				\hline
				50&1554&1590&31333&1580&22742\\	
				\hline								
				60&2126&2145&41557.2&2126&34603\\	
				\hline
				70&2340	&2369&49117.6&2367&36981\\	
				\hline	
				80&2575&2616&48363.8&2603&39327\\	
				\hline
				90&2884	&2909&66554.2&2897&40698\\	
				\hline
				100&2925&2946&61695.4&2938&48981\\	
				\hline
		\end{tabular}}
		\label{tab:resgwo}
	\end{center}
\end{table}

\begin{table}[h!]
	\caption{Wait metric in different methods }
	\begin{center}
		\scalebox{0.75}{
			\begin{tabular}{|c|c|c|c|c|c|}
				\hline
				\textbf{\# Task }&\textbf{Min-Based }&\textbf{Max-Based}&\textbf{Random}&\textbf{MET }&\textbf{GWO-Based}\\
				\hline
				10&131&	328	&7403.6&179&5732\\
				\hline
				20&242&565&11627.8&291&6299\\
				\hline
				30&253&846&	18134.2&406&13145\\
				\hline
				40&432&1320	&29063.4&623&17718\\	
				\hline
				50&514&	1422&30460.4&719&21856\\	
				\hline								
				60&671&1968&40557.8&929&33201\\	
				\hline
				70&800&2194&47960.6&1005&35809\\	
				\hline	
				80&810&2400&46963&1073&37767\\	
				\hline
				90&930&2699&59485.8&1214&39191\\	
				\hline
				100&931&2711&60142.2&1129&47558\\	
				\hline
		\end{tabular}}
		\label{tab:waitgwo}
	\end{center}
\end{table}

\textcolor{black}{In addition, the GWO-based proposed method is compared with the normal mode where the proposed method is executed without vehicular fog computing. It is supposed to utilize data center and static fog nodes and to remove dynamic fog nodes. The obtained results are reported in Table~\ref{tab:comparewithoutVFC}. As shown in Table~\ref{tab:comparewithoutVFC}, the proposed method based on VFC has shown better results than the method without VFC. Because numerous vehicles are employed as computing nodes in VFC, therefore; it can reduce delay and makespan.}

\begin{table}[h!]
	\caption{\textcolor{black}{Wait, makespan, and cost metric in the proposed method with/without dynamic fog nodes.} }
	\begin{center}
		\scalebox{0.75}{
			\begin{tabular}{|c|c|c|c|c|c|c|}
				\hline
				&\multicolumn{3}{|c|}{\textbf{GWO-based using cars }}&\multicolumn{3}{|c|}{\textbf{GWO-based without using cars}}\\
				\hline
				\textbf{\# Task }&\textbf{wait}&\textbf{makespan}&\textbf{cost}&\textbf{wait}&\textbf{makespan}&\textbf{cost}\\
				\hline
				10&5732&5970&10838&6285&6560.8&12454	\\
				\hline
				20&6299&6588&16917&8719&9251&22317\\
				\hline
				30&13145&13660&26810&13703&14744&32410\\
				\hline
				40&17718&18524&40633&22829&24061&48008\\	
				\hline
				50&21856&22742&44020&22605&24232&57884\\	
				\hline								
				60&33201&34603&61025&32083&33883&71315\\	
				\hline
				70&35809&36981&68070&42619&44534&79107\\	
				\hline	
				80&37767&39327&76220&42372&44491&85692\\	
				\hline
				90&39191&40698&84864&48676&50935&97322\\	
				\hline
				100&47558&48981&86178&51779&54240&98912\\	
				\hline
		\end{tabular}}
		\label{tab:comparewithoutVFC}
	\end{center}
\end{table}

According to both the supposed scenarios modifying the number of fog nodes and modifying coefficients for the objective function in Eq.~\ref{eq12}, the experiments were repeated and the results are reported in the following.

\textbf{Number of fog nodes:} In this experiment, we changed the number of fog nodes. The same tasks were handed over to the system for scheduling and execution. First, the number of static fog nodes is changed  and the number of dynamic fog nodes is constant. Second, the number of static fog nodes is fixed and in contrast number of dynamic fog nodes is changed. The number of static and dynamic fog nodes is increased and different metrics are computed. The results are shown in Table.~\ref{tab:mfgs} and Table.~\ref{tab:mfgd}. Table.~\ref{tab:mfgs} shows by increasing the number of static fog nodes, makespan and wait are improved, although, the cost metric is increased. Table.~\ref{tab:mfgd} indicates that increasing the number of dynamic fog nodes results in better makespan and wait. The cost value is improved by increasing iteration and dynamic fog nodes. The identical \textit{search agent} value (set to 70) has been considered for all experiments in both Tables.~\ref{tab:mfgs} and ~\ref{tab:mfgd}.

\begin{table}[h!]
	\caption{Makespan, wait, and cost metric in the proposed method using modifying static fog nodes}
	\begin{center}
		\scalebox{0.85}{
			\begin{tabular}{|c|c|c|c|c|c|}
				\hline
				\textbf{\# Task}&\textbf{Iteration}&\textbf{Makespan}&\textbf{Wait}&\textbf{Cost}&\textbf{Static fog nodes}\\
				\hline
				\multirow{4}{*}{50} &\multirow{2}{*}{50}&22742&21856&44020&5\\
				
				&&21567&20764&47468&7 \\\cline{3-6}
				&\multirow{2}{*}{100}&20831&20103&44288&5\\
				&&21169&20200&47187&7\\
				\hline
		\end{tabular}}
		\label{tab:mfgs}
	\end{center}
\end{table}
    
\begin{table}[h!]
	\caption{Makespan, wait, and cost metric in the proposed method using modifying dynamic fog nodes}
	\begin{center}
		\scalebox{0.85}{
			\begin{tabular}{|c|c|c|c|c|c|}
				\hline
				\textbf{Task}&\textbf{Iteration}&\textbf{Makespan}&\textbf{Wait}&\textbf{Cost}&\textbf{Dynamic fog nodes}\\
				\hline
				\multirow{4}{*}{50} &\multirow{2}{*}{50}&22742&21856&44020&10\\
				
				&&21863&21030&44752&12 \\\cline{3-6}
				&\multirow{2}{*}{100}&20831&20103&44288&10\\
				&&20502&19664&44100&12\\
				\hline
		\end{tabular}}
		\label{tab:mfgd}
	\end{center}
\end{table}

\textbf{Analysis of ($\alpha$, $\beta$)}: Two coefficients $\alpha$ and $\beta$ can be effective on the results based on Eq.~\ref{eq12}. In addition to ($\alpha$, $\beta$)=(0.5, 0.5), we consider two different states for ($\alpha$, $\beta$), including (0.25, 0.75), and (0.75, 0.25). In fact, we give more significance to cost or makespan using modifying coefficients in Eq.~\ref{eq12}. The results are acquired by supposing the number of tasks to be 50. Other important parameters in the GWO algorithm are investigated. 

\begin{table}[h!]
	\caption{Different metrics based on various values of \textit{iteration} at ($\alpha$, $\beta$)=(0.25, 0.75) }
	\begin{center}
		\scalebox{0.85}{
			\begin{tabular}{|c|c|c|c|c|}
				\hline
				\textbf{Iteration }&\textbf{Cost function }&\textbf{Cost}&\textbf{Makespan}&\textbf{Wait }\\
				\hline
				50&0.210709537&44376.2&19942.2&19060.4\\
				\hline
				100&0.210640167&44390.6&\textbf{19408.6}&\textbf{18580.2}\\
				\hline
				150&0.211985461&44081&21371&20521.4\\
				\hline
				200&\textbf{0.215883626}&\textbf{43298}&19983.8&19255.2\\	
				\hline
		\end{tabular}}
		\label{tab:res125}
	\end{center}
\end{table}

Table.~\ref{tab:res125} shows that with the number of iterations of 100, the time values are improved and the cost value is better as the number of repetitions increases. Table.~\ref{tab:res225} optimizes the makespan and wait metrics for the value of 2 for the $\vec{a}$ parameter. Further, by increasing the search agent, the results are improved as described in Table.~\ref{tab:res325}. It should be noticed that these findings are also demonstrated in all three Tables.~\ref{tab:res175}, ~\ref{tab:res275}, and ~\ref{tab:res375}. All findings in Table.~\ref{tab:res175} are similar to Table.~\ref{tab:res125}. For makespan and wait metrics, all best outcomes are for $\vec{a}$ equal to 2 in Table.~\ref{tab:res275}. Lastly, the best results for the search agent are equal to 70 in Table.~\ref{tab:res375} are achieved the same as Table.~\ref{tab:res325}.  

\begin{table}[h!]
	\caption{Different metrics based on various values of $\vec{a}$ at ($\alpha$, $\beta$)=(0.25, 0.75)}
	\begin{center}
		\scalebox{1}{
			\begin{tabular}{|c|c|c|c|c|}
				\hline
				\textbf{a }&\textbf{Cost function }&\textbf{Cost}&\textbf{Makespan}&\textbf{Wait }\\
				\hline
				3&\textbf{0.212887805}&\textbf{43861}&24242.8&23372\\
				\hline
				2&0.210640167&44390.6&\textbf{19408.6}&\textbf{18580.2}\\
				\hline
				1&0.212456844&44014.6&20665.4&19769.6\\
				\hline
				
		\end{tabular}}
		\label{tab:res225}
	\end{center}
\end{table}

\begin{table}[h!]
	\caption{Different metrics based on various values of \textit{search agent} parameter at ($\alpha$, $\beta$)=(0.25, 0.75)}
	\begin{center}
		\scalebox{0.85}{
			\begin{tabular}{|c|c|c|c|c|}
				\hline
				\textbf{Search agent }&\textbf{Cost function }&\textbf{Cost}&\textbf{Makespan}&\textbf{Wait }\\
				\hline
				4&0.199921703&46663&32350&	31395\\
				\hline
				5&0.200658739&46797.2&25150.4&24278.8\\
				\hline
				20&0.204870982&45585.4&24441.8&23547.2\\
				\hline
				50&\textbf{0.212891362}&\textbf{43904.6}&20602.4&19796\\
				\hline
				70&0.210640167&	44390.6&\textbf{19408.6}&\textbf{18580.2}\\
				
				\hline
		\end{tabular}}
		\label{tab:res325}
	\end{center}
\end{table}

\begin{table}[h!]
	\caption{Different metrics based on various values of \textit{iteration} at ($\alpha$, $\beta$)=(0.75, 0.25) }
	\begin{center}
		\scalebox{0.85}{
			\begin{tabular}{|c|c|c|c|c|}
				\hline
				\textbf{Iteration }&\textbf{Cost function }&\textbf{Cost}&\textbf{Makespan}&\textbf{Wait }\\
				\hline
				50&0.639928652&	43613.4&19564.4&18776.6\\
				\hline
				100&0.636595956&43869.2&\textbf{18846.8}&\textbf{18015.4}\\
				\hline
				150&0.631775457&44201.4&21414.6&20541.2\\
				\hline
				200&\textbf{0.645654497}&\textbf{43220}&21397.6&20549.4\\	
				\hline
		\end{tabular}}
		\label{tab:res175}
	\end{center}
\end{table}

\begin{table}[h!]
	\caption{Different metrics based on various values of $\vec{a}$ at ($\alpha$, $\beta$)=(0.75, 0.25)}
	\begin{center}
		\scalebox{1}{
			\begin{tabular}{|c|c|c|c|c|}
				\hline
				\textbf{a }&\textbf{Cost function }&\textbf{Cost}&\textbf{Makespan}&\textbf{Wait }\\
				\hline
				3&0.630331717&	44259.4&	21527.6	&20727.2\\
				\hline
				2&0.636595956&43869.2&\textbf{18846.8}&\textbf{18015.4}\\
				\hline
				1&\textbf{0.644276662}&\textbf{43302}&19613.6&18764.2\\
				\hline
				
		\end{tabular}}
		\label{tab:res275}
	\end{center}
\end{table}

\begin{table}[h!]
	\caption{Different metrics based on various values of \textit{search agent} parameter at ($\alpha$, $\beta$)=(0.75, 0.25)}
	\begin{center}
		\scalebox{0.85}{
			\begin{tabular}{|c|c|c|c|c|}
				\hline
				\textbf{Search agent }&\textbf{Cost function }&\textbf{Cost}&\textbf{Makespan}&\textbf{Wait }\\
				\hline
				4&0.598785832&46584.2&28319.2&27502.2\\
				\hline
				5&0.601458564&46374.2&25301.8&24407\\
				\hline
				20&0.624489354&44671.4&20276.8&19366.2\\
				\hline
				50&0.631176733&	44191.4&21903.4&21018\\
				\hline
				70&\textbf{0.636595956}&	\textbf{43869.2}&\textbf{18846.8}&\textbf{18015.4}\\
				
				\hline
		\end{tabular}}
		\label{tab:res375}
	\end{center}
\end{table}

\subsection{Scientific workflow evaluation}
\label{s54}
To further validate the proposed method, in addition to selecting random tasks, we also considered an application for analyzing. There are many different types of applications such as compute-intensive, data-intensive, batch processing, scientific workflows, bag of tasks, and image processing applications. Scientific applications are usually in the form of a bag of task models composed of independent tasks. Various problems in different fields such as massive searches (such as key breaking), computational biology, tumor detection, and image processing are modeled and implemented as a set of tasks. According to research findings, between 34 and 89 percent of the tasks accepted for parallel systems are from the bag of tasks model \cite{ minh2013parallel}. For this purpose, we explore a scientific workflow for precisely evaluating the task scheduling problem based on the proposed method. This scientific application includes a set of independent tasks in its workflow and also there is more than one task at one level. At first, a brief description of the scientific workflow is presented.

The Montage astronomy workflow has commonly been employed extensively for evaluating workflow algorithms and systems \cite{juve2013characterizing}. Montage, an open-source toolkit developed by the NASA/IPAC Infrared Science Archive, allows users to construct customized mosaics of the sky using input images in the Flexible Image Transport System (FITS) format. The geometry of the output image is determined by calculating it from the input images during the creation of the final mosaics. The inputs are subsequently transformed to ensure they have the same spatial size and rotation. The background emissions in the images are adjusted to achieve a consistent level, and the transformed, adjusted images are combined to create the final mosaic output. 

\begin{figure}[h]
	\begin{center}
		\includegraphics[scale=0.65]{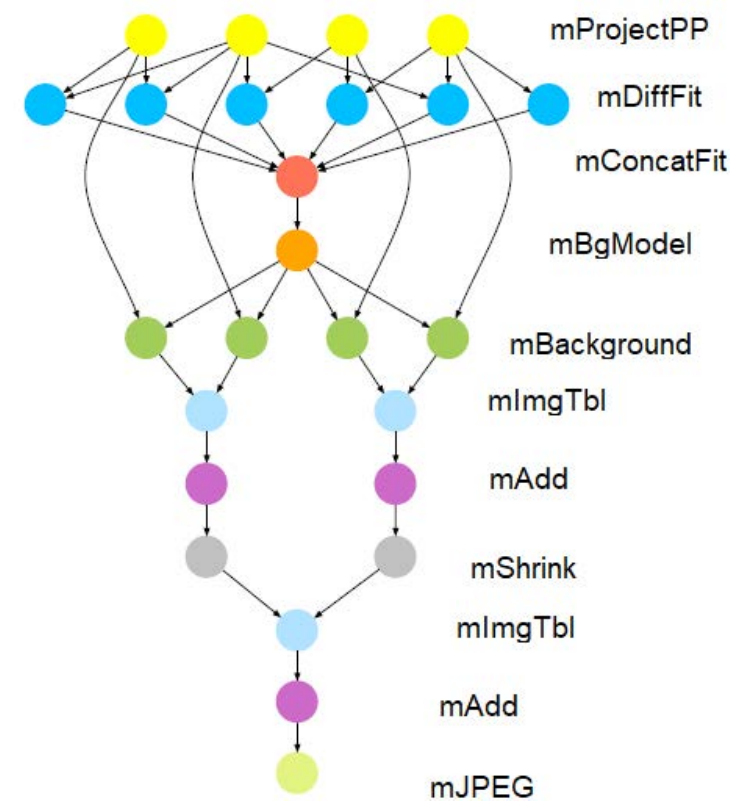}
		\caption{Workflow of Montage \cite{juve2013characterizing}}
		\label{Mont}
	\end{center}
\end{figure}

Figure.~\ref{Mont} illustrates the workflow of the Montage application. The obligations of the job include: (1). mProjectPP, (2). mDiffFit, (3). mContactFit, (4). mBgModel, (5). mBackground, (6). mImgTbl, (7). mAdd, (8). mShrink , and (9). mJPEG.         

As you can observe from the figure, in each Montage application, there is a mConcatFit, mBgModel and mJPEG task. Their execution time is 143, 384, and 0.64 seconds, respectively. And rest consists of 16, 17, 2102, and 6172 tasks. In total, the number of its tasks is equal to 10429 by considering all types of tasks. The execution time of different tasks is shown in Table.~\ref{tab:mont}. Interested readers can refer to \cite{juve2013characterizing} for more information about the Montage workflow.

\begin{table}[h!]
	\caption{Example of execution of Montage \cite{juve2013characterizing} }
	\begin{center}
		\scalebox{0.85}{
			\begin{tabular}{|c|c|c|}
				\hline
				\textbf{Task }&\textbf{Number}&\textbf{Execution time (in second)}\\
				\hline
				1&2102&1.73\\
				\hline
				2&6172&0.66\\
				\hline
				3&1&143.26\\
				\hline
				4&1&384.49\\
				\hline
				5&2102&1.72\\
				\hline
				6&17&2.78\\
				\hline
				7&17&282.37\\
				\hline
				8&16&66.10\\
				\hline
				9&1&0.64\\
				\hline
		\end{tabular}}
		\label{tab:mont}
	\end{center}
\end{table}

\begin{figure}[h!]
	\begin{center}
		\includegraphics[scale=0.5]{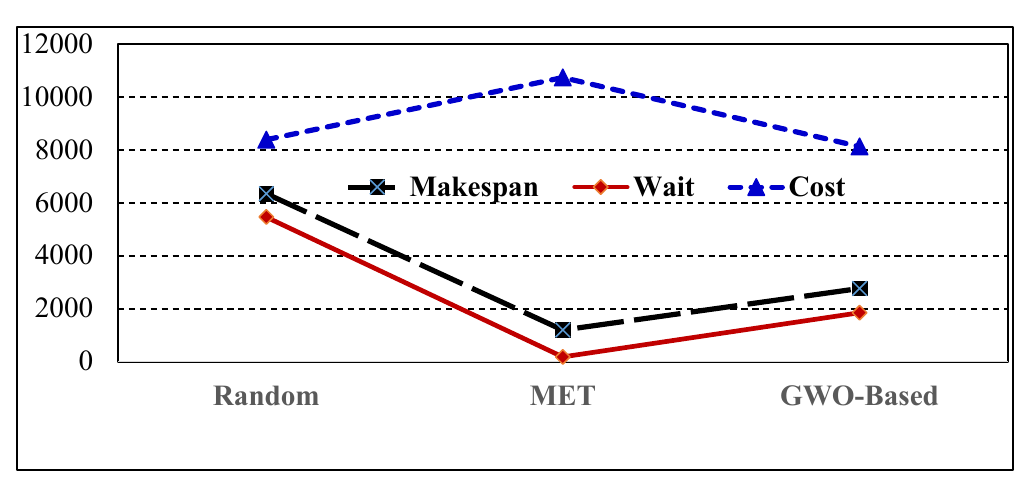}
		\caption{Comparison of Montage in three methods. }
		\label{ResMont}
	\end{center}
\end{figure}

The Montage was evaluated by the proposed method and compared with MET and random method. The results for cost, makespan, and wait are drawn in Figure.~\ref{ResMont}.

The results verify that our method based on GWO achieves the best result in cost compared to random and MET methods, although wait and makespan metrics have shown a worse amount than MET. {In overall, we presented a GWO-based method for the VFC problem with aiming to decrease monetary cost. Our findings verify that the proposed method is able to reasonably decrease cost for both randomly generated tasks and real applications.}

\subsection{Discussion}
\label{s55}
\textcolor{black}{In total, the VFCs play an important role in ITS. They can utilize the power of vehicular processors in different areas in order to perform many applications such as real-time requests, entertainment, and virtual reality. In addition to VFC advantages, there are challenges that it is essential to examine them. It is important to notice that in VFC systems, it is possible to fail a dynamic fog node or move a vehicular and leave a specific area when a fog node is running a task. There are approaches to resolving this problem. For example, tasks can be migrated to cloud data centers, and offloaded to other vehicles in the same area \cite{mishra2023collaborative}. All computation and saved data are transmitted to the alternate computing nodes. Task distribution across multiple fog nodes is suggested to reduce the impact of individual node failures. Likewise, mechanisms can be implemented to periodically back up data to a central server or cloud storage. Network management can be performed continuously and it monitors the network for node failures, link failures, and congestion. Finally, these strategies can be combined and as a result, VFC systems can mitigate the risks associated with node failures and disconnections, and ensure the continuity of computing operations and the integrity of data.}

\textcolor{black}{To sum up, GWO's simplicity, efficiency, exploration-exploitation balance, computational efficiency, and scalability make it a promising method for task scheduling in vehicular fog computing systems. }

\section{Conclusion}
\label{}

This paper investigated the task scheduling problem based on vehicular fog computing systems in a fog-cloud environment. The proposed algorithm is based on grey wolf optimization and allocates resources into tasks. In this study, static fog nodes being given priority over dynamic nodes are examined for the scheduling process. Nodes with more computing resources also have priority. The objective function follows decreasing cost and time. The experimental results showed that the proposed algorithm acquires desirable cost in compared to allocation methods such as random, maximum, and minimum. Furthermore, we modified the number of static and dynamic fog nodes. Results verified that the number of dynamic fog nodes can be effective in achieving better results. In future work, we will suppose mobility in dynamic fog nodes and investigate the effectiveness of the proposed method.      






\bibliographystyle{elsarticle-num} 
\bibliography{GWOA}
%
%
%
\end{document}